# Enhanced polarization, magnetoelectric response and possible bi-multiferroic phase separation in $Tb_{1-x}Ho_xMnO_3$

C. L. Lu, S. Dong, K. F. Wang, and J. –M. Liu[*]

*Laboratory of Solid State Microstructures, Nanjing University, Nanjing 210093, China*

*International Center for Materials Physics, Chinese Academy of Science, Shenyang, 110016, China*

**[Abstract]** A series of manganites $Tb_{1-x}Ho_xMnO_3$ ($0 \leq x \leq 0.6$) with orthorhombic structure are synthesized and detailed investigations on their multiferroicity are performed. Successive magnetic transitions upon temperature variation are evidenced for all samples, and both the $Mn^{3+}$ spiral spin ordering and rare-earth spin ordering are suppressed with increasing $x$. Significant enhancement of both the polarization and magnetoelectric response within $0.2 < x < 0.4$ is observed, which may be ascribed to the competition possibly existing between spiral and $E$-type spin orders. Theoretical calculation is given based on two $e_g$-orbital double-exchange model, and the result supports the scenario of the multiferroic phase separation.

---

[*]Correspondent, E-mail: liujm@nju.edu.cn

# I. Introduction

Multiferroic manganites $R$MnO$_3$ ($R$=Tb, Dy etc) with orthorhombic structure have been intensively investigated since the discovery of gigantic magnetoelectric (ME) effect in TbMnO$_3$.[1-4] They show simultaneous reversal of ferroelectric (FE) polarization ($P$) and magnetization ($M$) upon sweeping magnetic field $H$.[5] This cross-coupling between $M$ and $P$ is of fundamental interest and provides an additional degree of freedom in designing memory elements. Investigations [6-12] revealed that the noncollinear spiral spin order is essential to onset of polarization. A possible microscopic mechanism of the spiral-spin driven ferroelectricity is the inverse Dzialoshinski-Moriya (DM) interaction or spin current scenario,[7,13,14] in which two nearest-neighbor noncollinear spins displace the intervening oxygen through the electron-lattice interaction. Accordingly, the correlation between $P$ and two noncollinearly coupled spins is given by $\boldsymbol{P} \propto \boldsymbol{e}_{ij} \times (\boldsymbol{S}_i \times \boldsymbol{S}_j)$, where $\boldsymbol{e}_{ij}$ denotes the unit vector connecting the two spins $\boldsymbol{S}_i$ and $\boldsymbol{S}_j$.

However, when $A$-site ion Tb$^{3+}$ (Dy$^{3+}$) is replaced by Ho$^{3+}$ which has smaller ionic radius,[15] the Mn spins form a collinear $E$-type antiferromagnetic ($E$-AFM) order with commensurate wavevector [16] and it was predicted [17] that FE order can exist in this $E$-AFM structure, caused by a balance between double-exchange (DE) electron-hopping and elastic energy, and that the quantum-mechanical effects of electron-orbital polarization is vital to ferroelectricity.[18] These predictions were evidenced in HoMnO$_3$ recently.[19] Here, one notes that the origin of ferroelectricity in $E$-AFM magnetic structure is different from that in spiral-spin order (SSO) systems, and a first order phase transition between SSO and $E$-AFM has been predicted.[9] Along this line, a natural question arises: what is the ground state given the possible competition between the SSO and $E$-AFM orders, especially when A-site disorder presents, and how is the coupling between $P$ and $M$ near the phase boundary between the two types of spin orders, referring to the colossal magnetoresistance (CMR) effect observed in multi-order competing manganites.[20] A scheme of the possible bi-multiferroic phase separation (PS) is presented in Fig. 1, and the possible PS region is indicated by the red shadow. Indeed, a magnetic phase separation state was evidenced in the phase boundary between the A-type AFM order and SSO. [21]

In this article, we address this issue by focusing on TbMnO$_3$ as the prototype sample via successive doping of $Tb^{3+}$ with $Ho^{3+}$, intending to reach a state with competing spiral-spin and $E$-AFM orders in $Tb_{1-x}Ho_xMnO_3$. Subsequently, we perform a series of measurements on the multiferroicity. We will reveal the enhanced $P$ and ME response between $P$ and $M$, which is argued to be due to the competition probably existing between the SSO and $E$-AFM. This $Ho^{3+}$-substitution effect is further studied by a theoretical calculation, which reveals consistency with our experimental results. The remainder of this paper is organized as follow. We report the sample preparation and property characterizations using various techniques in Sec.II. The details of the measured data are presented in Sec.III, where we discuss every aspect of the doping effects in $Tb_{1-x}Ho_xMnO_3$ ($0 \leq x \leq 0.6$). Theoretical calculation and comparison with the experiments are presented in Sec.IV. A short summary is given in Sec.V.

## II. Experimental procedure

Polycrystalline $Tb_{1-x}Ho_xMnO_3$ ($0 \leq x \leq 0.6$) were prepared by conventional solid-state reaction. The highly purified powders of oxides were mixed in stoichiometric ratios, ground, and then fired at 1200°C for 24 hours (h) in an oxygen flow. The resultant powders were reground and pelletized and then sintered at 1350°C for 24 hours in an oxygen flow with intermediate grindings. X-ray diffraction (XRD) with Cu $K\alpha$ radiation was performed on these samples at room temperature ($T$). The magnetic and specific-heat measurements using SQUID and PPMS were performed to probe the spin orders. For measuring dielectric constant $\varepsilon$ and $P$, gold pastes were used as electrodes and varying $T$ and $H$ environment was provided by PPMS. We measured $\varepsilon$ using the HP4294 impedance analyzer, while $P$ as a function of $T$ was evaluated from the pyroelectric current measured by the Keithley 6514A electrometer after cooling the samples under selected electric fields, and detailed discussion on the state-of-merit of this measurement can be found in earlier report.[19,21]

## III. Results and discussion

### A. Multiferroic behaviors

We identify the crystallinity and structure of the as-prepared samples. The XRD patterns of all samples are presented in Fig.2. All the reflections can be indexed with a single orthorhombic structure of space group Pbnm. A continuous shifting towards high 2θ side with increasing x, as more clearly identified in the inset, is observed, indicating the slight lattice contraction with increasing x. This variation is reasonable because of slightly smaller $Ho^{3+}$ than $Tb^{3+}$. We pay more attention to the magnetic ordering and corresponding P-generation. As an example, we present in Fig.3(a) and (b) the measured M, specific-heat divided by T (C/T), ε, and P as a function of T for sample x=0 ($TbMnO_3$). It is clearly shown that the measured M is small and shows no anomaly over the whole T-range, indicating no ferromagnetic spin correlation. The reason lies in the fact that all possible spin orders are antiferromagnetic-type.[22] We then consult to the C/T data which is sensitive to such types of ordering. The anomaly at T=$T_N$=40K corresponds to the onset of a collinear sinusoidal spin-ordering with an incommensurate wavevector, where ε(T) also exhibits a distinct kink. A second anomaly at T=$T_C$=25K develops a noncollinear spiral spin order with a T-independent wavevector, at which nonzero P ensues and is accompanied with a sharp peak in ε(T). Upon further cooling down to T=6K, a third anomaly of C/T associated with the long-range ordering of $Tb^{3+}$ moments can be identified. These one-to-one correspondences between the $Mn^{3+}$ spin orders and polarization generation do indicate the strong cross-coupling between magnetic and ferroelectric orders.

Now we address the evolution of the measured data with varying *x*, and three typical sets of data are shown in Fig.3(c) and (d), respectively. It seems that $T_N$ is independent of *x*, but $T_C$ and $T_R$ show continuous downshifting as *x* increases from 0.0 to 0.5. Considering the facts that $T_N$ for both $TbMnO_3$ and $HoMnO_3$ are ~41 K and no much additional lattice distortion associated with the doping is generated since the $Tb^{3+}$ radius (~1.095Å) is close to $Ho^{3+}$ radius (~1.072 Å),[15] one expects no remarkable change of $T_N$ upon the doping, a reasonable argument and consistent with measured results. However, because $T_C$ is the onset point of spiral spin order, $Ho^{3+}$-doping at $Tb^{3+}$ site indicates the suppression of the spiral spin ordering down to a lower *T*, characterized by downshift of this onset point. It is also identified that $T_R$ disappears at *x*~0.5, revealing no more $Tb^{3+}$ long-range spin order possible at the half-doping. What is most interesting here is the *x*-dependence of the measured P at low *T*. It is seen that the value of P for sample *x*=0.3 is almost double that for $TbMnO_3$, indicating significant enhancement of P upon such a doping. Surely, for an over-high doping, e.g. *x*=0.5 here, we observe downshift of $P$ back to a value similar to HoMnO$_3$.

To obtain a more comprehensive picture of the doping effect, $P$ as a function of poling field $E$ at the lowest $T=2$ K for selected samples are evaluated, and the data for samples $x=0$ and 0.3 is shown in Fig.4(a). It reveals that the remnant polarization $P_r$ can be roughly obtained at $E>5$ kV/cm. Clearly, sample $x=0.3$ possesses a much larger $P_r$ than sample $x=0$, indicating significantly enhanced ferroelectricity with respect to sample $x=0.0$. In Fig. 4(b), the measured $T_C(x)$ and $P_r(x)$ at $E=8$kV/cm with more data (phase diagram) is presented. We observe a roughly linear suppression of $T_C$ with $x$ until $x=0.5$ at which $T_C$ reaches its minimal of ~19.5K, followed by a slight return at $x=0.6$ beyond which the sample synthesis by conventional route becomes tough. Such a suppression can be understood by the doping induced $E$-type spin order (in HoMnO$_3$) in coexistence with the spiral spin order in TbMnO$_3$. This fact may allow one to expect a smooth suppression of $P_r$ in term of both the onset point ($T_C$) and magnitude, upon the doping. Nevertheless, the multiferroic physics associated with the Ho$^{3+}$-doping seems more complicated. The measured $P_r(x)$ at $T=2$K exhibits very unusual behavior, as shown in Fig. 4(b) too. Over range $0<x<0.6$, $P_r(x)$ does not decay in synchronization with $T_C(x)$. Instead, it does not change much with $x$ up to ~0.2 at which $P_r(x)$ exhibits a jump from 87 μC/m$^2$ to ~210 μC/m$^2$, a significant enhancement. This enhancement is maintained until $x\sim 0.4$ at which $P_r(x)$ again falls down to ~110μC/m$^2$ and then decays slowly with $x$.

### B. Magnetoelectric response

In order to get a deeper understanding of this doping effect, we measure $P(T)$ at several selected $H$ for selected samples, as shown in Fig.5. For sample $x=0$, $T_C$ is independent of $H$, consistent with earlier report,[1] a small increase of $P$ below $T_C$ with increasing $H$ is identified, associated with the multiple domains. For the doped samples, $P$ can be significantly reduced by $H$, and the largest ME response (defined as ME=$(P(0)-P(H))/P(0)$) is evidenced for sample $x=0.3$ and reaches about ~86% at $H=9$T at low $T$. It should be specially addressed that such a large ME response can't be obtained at $x<0.2$ and $x>0.4$, indicating the one-to-one correspondence between the $P$-enhancement and ME response. On the other hand, the dielectric susceptibility also shows unusual response to the magnetic field for sample $x=0.3$, and ε evolves with $T$ in a different way when $x>0.3$, as shown in Fig. 6. Considering the fact that the $E$-AFM spin order (HoMnO$_3$) contributes a lot to the dielectric constant, but a spiral one (TbMnO$_3$) does not.[19] This suggests that the $E$-type phase dominated within samples $x>0.3$, and sample $x=0.3$ after the magnetic field applied, which is in agreement with the above results.

### C. Discussion

An understanding of the physics underlying the Ho$^{3+}$-doping effects is challenging. For the unusual enhancement of $P_r(x)$ over $x\in(0.2, 0.4)$: (1) the large enhancement of $P_r$ can be simply ascribed to neither the contribution of the spiral spin order (favored in TbMnO$_3$) nor the $E$-type spin order (favored in HoMnO$_3$), since the $P$-value of either polycrystalline TbMnO$_3$ (~87 μC/m$^2$) or orthorhombic HoMnO$_3$ (~90 μC/m$^2$), is much smaller than ~210 μC/m$^2$. (2) If the Ho$^{3+}$-doping causes simply the solid-solution effect that the magnetic configuration is composed of spiral spin order as in TbMnO$_3$ and $E$-type spin order as in HoMnO$_3$, $P$ is expected to be within [87, 90] μC/m$^2$. Referring to the CMR manganites, where multi-order competing phases exist and the true CMR effects ensues at the phase boundary,[20] one possible explanation is the competition between spiral and $E$-type spin orders, since the two essentially different phases are neighboring to each other in the phase diagram. Indeed, a first-order transition was predicted theoretically between the two essentially different spin orders, and it is reasonable to anticipate such a two-phase competing in energy landscape, and thus the two phase coexistence.[9] Surely, we don't have sufficient evidence with such a competition, which appeals for additional exploration.

On the other hand, for the large ME response in sample $x=0.3$. We note that in TbMnO$_3$, an applied $H$ rotates the spiral plane from the $bc$ plane to the $ab$ plane, and the corresponding $P$ is switched from the $c$-axis to the $a$-axis.[4] This spiral plane rotation will not induce significant variation of $P$ in a polycrystalline sample, as shown in Fig.5(a) for sample $x=0$. While for the $E$-AFM spin order induced ferroelectricity, external $H$ suppresses the $E$-type order rather than enhancing it, and large ME response has been evidenced.[19] So one may correlate the large ME effects in Tb$_{1-x}$Ho$_x$MnO$_3$ with the suppression of $E$-type spin orders. If this would be true, one should expect a monotonous increase of the ME response with $x$ and the maximum at $x=1$ (HoMnO$_3$). However, our results confirms that the ME response does not monotonically increase with $x$ but reaches the maximum around $x\sim0.3$, coincident with the $P_r(x)$ dependence. Also the ground state of sample $x=0.3$ can not be the $E$-AFM orders since its physical property is rather different from HoMnO$_3$. Therefore, the large ME response in our samples, for example sample x=0.3, is not only contributed from the rotation of spiral-plane or the suppression of $E$-AFM orders, but also from additional ingredient due to the competition possibly existing between spiral and $E$-AFM spin orders, which can be modulated by magnetic field, similar to the observation in multiferroics TbMn$_2$O$_5$.[23] Nevertheless, it should be mentioned that a clear understanding of the Ho$^{3+}$-doping effects revealed here still need theoretical support.

### IV. Theoretical study

To further understand above experimental results, and also to predict the possible bi-multiferroic PS in Tb$_{1-x}$Ho$_x$MnO$_3$ system, a two-orbital DE model simulation is performed. The model used here is almost the same with the original one in Ref. [9], which can reproduce the phase diagram of $R$MnO$_3$ by tuning the nearest-neighbor (NN) superexchange (SE) coefficient $J_{AF}$ and the next-nearest-neighbor (NNN) SE coefficient along $b$-axis $J_{2b}$. Considering the orthorhombic lattice distortion, the NNN SE along $a$-axis, which is weaker than $J_{2b}$, is neglected for simplication.[9] By increasing ($J_{AF}$, $J_{2b}$), the ground state can change from the A-AFM to SSO, and finally to $E$-AFM. For example, with a proper Jahn-Teller (JT) distortion $\lambda|Q_2|=1.5$, ($J_{AF}=0.087$, $J_{2b}=0.008$) gives the $k=1/6$ ($k$ is the wave number along the pesudocubic [100] and [010] direction) SSO phase while ($J_{AF}=0.1$, $J_{2b}=0.01$) corresponds to $E$-AFM phase, where the unit is the DE hopping $t_0\sim0.2$-$0.3$ eV. More details of our model and method can be found in Ref. [9].

A 12×12 lattice with periodic boundary conditions will be used as in Ref. [9]. In real SSO $R$MnO$_3$, e.g. Tb$_{1-x}$Dy$_x$MnO$_3$, the wave number $k$ can change continuously with $x$. However, the lattice here can only accommodate with those SSO phases which's periods are divisors of 12. In other words, the wave number $k$'s of possible SSO phase are discrete here: 1/12, 1/6, 1/4, 1/3, 1/2, which is limited by the finite size lattice. Larger lattice size can improve the precision of $k$. However, the computational CPU time will also increase extremely with the increasing lattice size when using the DE model. Although the classical spin models can handle larger lattices for the SSO phase [24], it is under debate whether the E-AFM order can exist in these models. Therefore, in the current stage, a 12×12 lattice for the DE model is the best choice.

In most previous theoretical studies on the A-site disorder in doped manganites, a random onsite potential field is applied, which originates from the alloy-mixed cations with different valences. While in the $Tb_{1-x}Ho_xMnO_3$ case here, Tb and Ho are both +3. The A-site disorder comes from the cation size difference only which affects the Mn-O-Mn bond-angles, thus the exchange interactions. To mimic this effect, a random field is added into the SEs while others (DE and JT distortion) are assumed to be unchanged for simplicity. In practice, as in alloy-mixed $Tb_{1-x}Ho_xMnO_3$, a random distribution of A-site cations are firstly generated on a 12x12x2 three-dimensional lattice, where 2 is for the two NN AO layers of $MnO_2$ plane under studied). Then the SEs are calculated based on the A-site cations' distribution. Each $J_{AF}$ is determined by its four NN A-site cations while each $J_{2b}$ is determined by its two NN A-site cation:

$$J_{AFi} = (n_i^{Tb} J_{AF}^{Tb} + n_i^{Ho} J_{AF}^{Ho})/4$$
$$J_{2bj} = (n_j^{Tb} J_{AF}^{Tb} + n_j^{Ho} J_{AF}^{Ho})/2 \qquad (1)$$

where $n_i$ denote the number of NN A-site cations (Tb or Ho) surrounding the bond $i$. In real $TbMnO_3$, the SSO wavenumber $k\approx0.14$ which correponds ($J_{AF}$=0.086, $J_{2b}$=0.006) in our model when the JT distortion $\lambda|Q_2|$ is fixed as 1.5. However, in the 12×12 lattice, this coefficient group will give a $k$=1/6 SSO phase due to the aforementioned finite size limit. Noting the k≥1/4 SSO is unavaible because its energy is always higher than that of $E$-AFM, the $k$=1/12 spiral has to be adopted as the start point to illustrate how $k$ increases with $x$. Thus, ($J_{AF}^{Tb}$, $J_{2b}^{Tb}$) is set as (0.085, 0.005) in the following simulation, which gives $k$=0.12 SSO in the infinite lattice and $k$=1/12 in the 12×12 lattice. ($J_{AF}^{Ho}$, $J_{2b}^{Ho}$) is set as (0.1, 0.01) which can stabilize a $E$-AFM spin order.

Initialized with a random spins pattern and fixed JT distortion, a Monte Carlo (MC) simulation is performed firstly at $T$=0.002 (~5-7 K) which is low enough to approach the ground state. After the MC simulation (12000 MC steps), a zero-$T$ optimization is used to improve the precision of ground state. For each concentration $x$, three independent random fields have been tested and the results are averaged. To reveal the evolution of spin orders with $x$, the spin structure factors (SSF) are calculated.. In addition, the ferroelectricity based on the DM mechanism ($\sim \sum_{<ij>} \mathbf{e}_{ij} \times (\mathbf{S}_i \times \mathbf{S}_j)$) and DE mechanism[25] are evaluated respectively.

The SSF values of characteristic wavevectors ($S(k, k)$) are shown in Fig. 7(a). With the increase of $x$, the $k=1/12$ SSO is rapidly suppressed while the $k=1/6$ spiral becomes the dominant one, as shown in Fig. 7(a). At $x=0.1$, the coexistence of $k=1/12$ and $k=1/6$ components and their large fluctuations suggest a PS. However, this phase separation is caused by the finite size limit which does not allow the continous modulation from $k=1/12$ to $k=1/6$, while in real materials this transition should be 2$^{nd}$ order. Meanwhile, the DM induced FE $P$ is enhanced by increasing $k$, as shown in Fig. 7[b]. Around $x=0.5$, a real PS occurs: the coexistence of $E$-AFM ($k=1/4$) and $k=1/6$ spiral according to the SSF result and FE $P$. Beyond $x=0.5$, $E$-AFM spin order becomes the dominant one and the FE $P$ is mainly driven by the DE mechanism. It should be pointed out that it is not meaningful to comparing the $x$ value to experimental one directly, because of the finite size limit as stated before.

The examples of optimized spin pattern, as shown in Fig. 8, also confirm the phase transitions with $x$. For (a), (b), (d), the spins' orders are almost pure $k=1/12$ ($x=0$), $k=1/6$ ($x=0.2$) spiral and $E$-AFM ($x=1$) respectively, while the $E$-AFM zigzag segments and SSO clusters coexist in (c) at $x=0.5$. In the PS region, the intensive phase competition and large fluctuation will certainly result in a sensitive response to external stimulates, such as enhanced magnetoelectric response.

In short, a DE model has been studied to mimic the Ho-substitution effects in $Tb_{1-x}Ho_xMnO_3$. Although limited by the finite size effects, a qualitative agreement between model simulation and experimental measurements remains robust. With increasing $x$, the FE $P$ is firstly enhanced by the increase of spiral wave nubmer $k$ (shortening of spiral period). Then near the SSO-E-AFM phase boundary, a bi-multiferroic PS occurs due to the A-site disorder. In this PS region, the ME response is more sensitive than the pure E-AFM and SSO phases. What should be mentioned is the difference between the multiferroic PS atate and CMR PS state in manganites. For the latter case, magnetic field always suppresses the charge-ordered phase but favors the ferromagnetic phase, giving rise to colossal response of the resistivity and magnetism. For the multiferroic manganites, the magnetic field would suppress the polarization in both the SSO phase and the E-AFM phase. [1, 19] In this sense, the multiferroic PS state we propose here would be much more complicated. The competition between the SSO and E-AFM order makes both the two ordered phases rather susceptible to magnetic field, and the ME effect can't be simply regarded as the resultant of suppressing one phase but favoring the other one. This issue needs more extensive investigation using advanced and specific techniques.

## V. Conclusion

In conclusion, we performed experimental and theoretical investigations on the magnetism, specific heat, electric polarization, and dielectric susceptibility in multiferroics manganites $Tb_{1-x}Ho_xMnO_3$. The appropriate $A$-site doping using $Ho^{3+}$ to replace $Tb^{3+}$ simultaneously enhances the electric polarization and magnetoelectric response significantly over the doping range from $x=0.2$ to $x=0.4$. We propose that there possibly exists a competition between the spiral spin order and E-type AFM order within this doping range. Our calculation shows good consistency with the experimental results in two aspects: (1) the electric polarization increases with increasing x and (2) a coexisting SSO phase and E-AFM phase (phase separation) near the phase boundary is identified. However, further microscopic investigations are appealed to directly reveal the possible multiferroic phase separation.

**Acknowledgement**

S.D. thanks E. Dagotto, S.-W. Cheong, T. Kimura and N. Nagaosa for helpful discussions. This work was supported by the Natural Science Foundation of China (50601013, 50832002), the National Key Projects for Basic Research of China (2009CB929501, 2006CB921802), and the 111 Programme of MOE of China (B07026).

*Figure Captions:*

Figure 1. Scheme of the possible multiferroic phase separation between spiral spin order (left upper panel) and *E*-type spin order (right upper panel). The red shadow in the bottom panel indicates the phase separation region.

Figure 2. Room temperature X-ray diffraction patterns of $Tb_{1-x}Ho_xMnO_3$ ($0\leq x\leq 0.6$). The inset shows the diffraction peaks of (112) and (200) planes for samples $x=0$, 0.1, 0.3, 0.5 from bottom to up, respectively.

Figure 3. Temperature profiles of specific heat divided by temperature *C/T* (a), dielectric constant $\varepsilon$ and electric polarization *P* (b), for sample $TbMnO_3$; (c) *C/T* and (d) *P* as a function of *T* for samples $x=0.1$, 0.3, 0.5, respectively. The *vertical dot lines* in (a) and (b) mark the $T_N$, $T_C$, $T_R$, respectively. The poling field is 3 kV/cm.

Figure 4. (a) *E*-dependence of *P* at *T*=2K for samples $x=0$ and 0.3. (b) Phase diagram of $Tb_{1-x}Ho_xMnO_3$ ($0\leq x\leq 0.6$) as a function of *x*. The red dots represent the paraelectric-ferroelectric transition point $T_C$, and the blue squares (*E*=8kV/cm) represent the measured $P_r$ at *T*=2 K. Both the red and blue lines are guide for eyes;

Figure 5. Magnetic field modulation of polarization as a function of *T* for samples $x=0$, 0.2, 0.3, 0.4, respectively. The poling field is 3 kV/cm.

Figure 6. Magnetic field modulation of dielectric constant as a function of *T* for sample x=0.2, 0.3, 0.4, respectively.

Figure 7. $Ho^{3+}$-doping level *x* dependence of (a) spin structure factors and (b) ferroelectric polarizations originated from the DM interaction and DE process. The phase separation between *E*-AFM and SSO phase occurs at *x*=0.5, indicated by coexistence of spin orders and large fluctuations (error bars). In (a), *k*=1/12, *k*=1/6 are for the SSO phases while *k*=1/4 is for the E-AFM phase. Both two polarizations are normalized to their saturation values (*k*=1/6 and

E-AFM) respectively. Since the real ratio between these two polarizations is unclear, the absolute value of total polarization is unavailable in our simulation.

Figure 8. Spin patterns of Mn cations obtained after the MC simulation and optimization. (a-d) are for $x$=0, 0.2, 0.5, 1, respectively. For (a), (b), (d), the spins' orders are almost pure $k$=1/12, $k$=1/6 spiral-spin orders and $E$-AFM respectively. The coexistence of SSO and $E$-AFM segments are highlighted in (c).

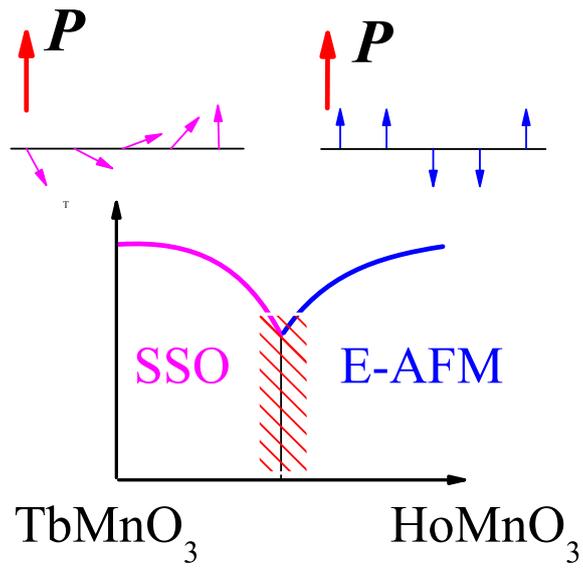

Figure 1

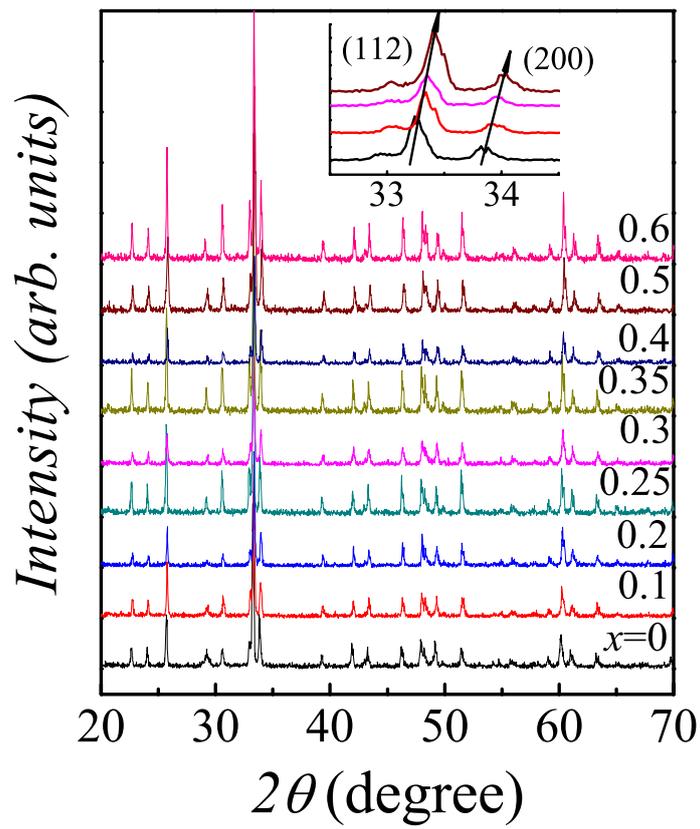

Figure 2

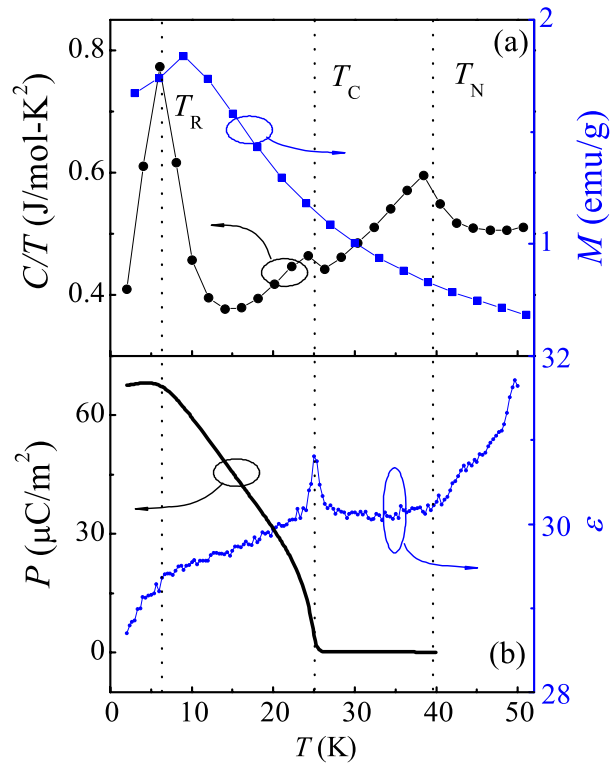

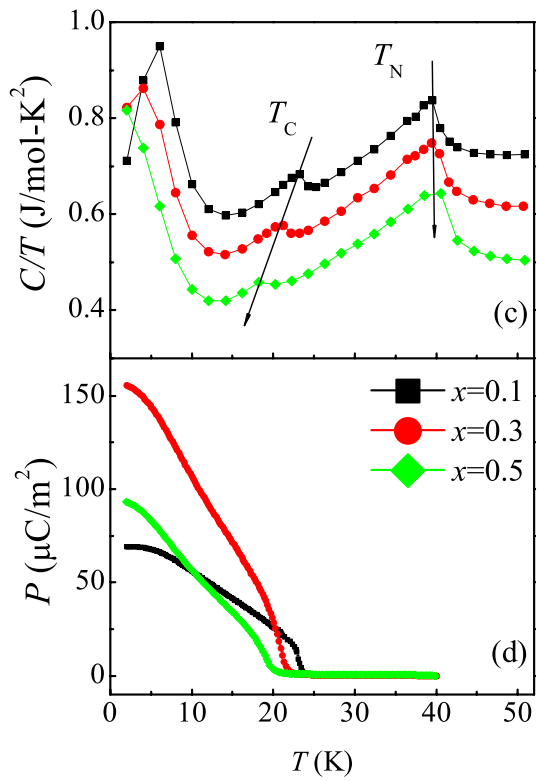

Figure 3

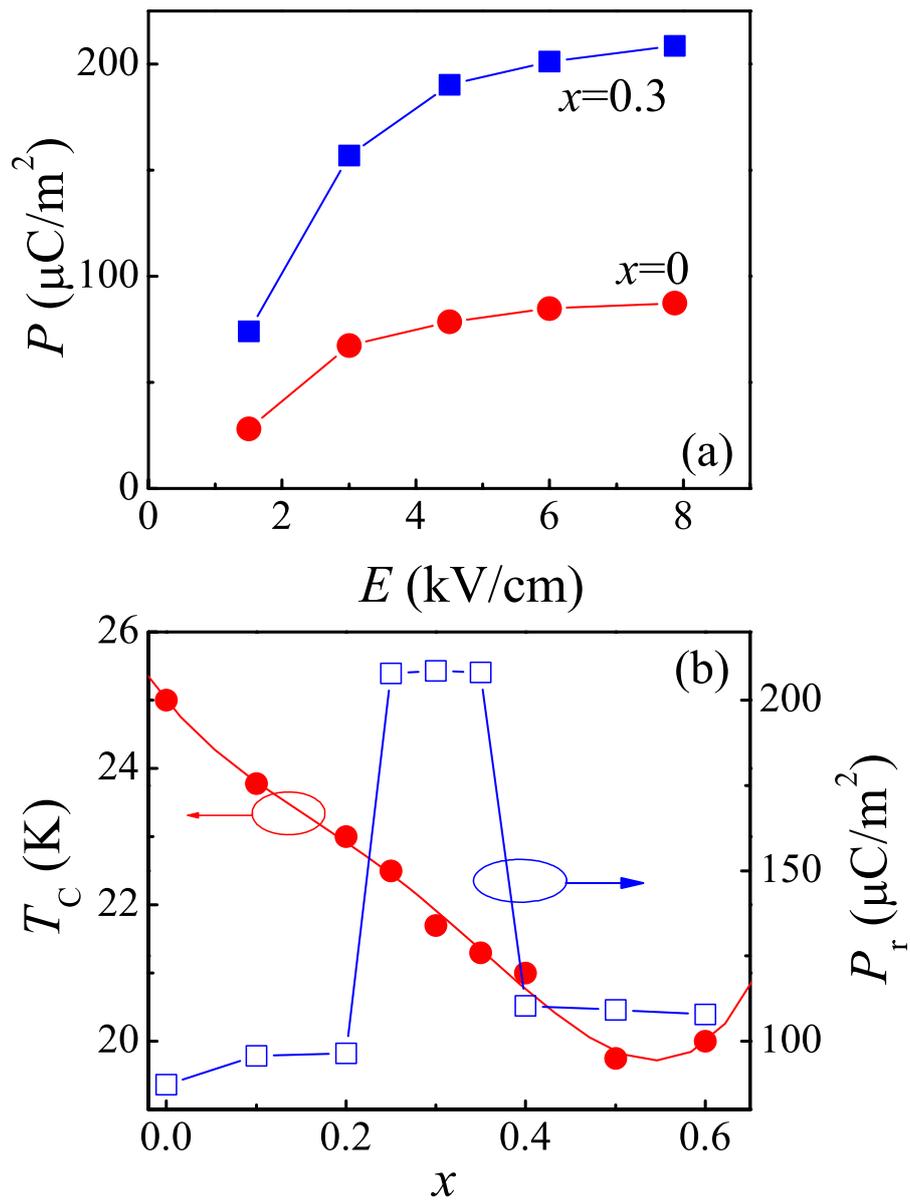

Figure 4

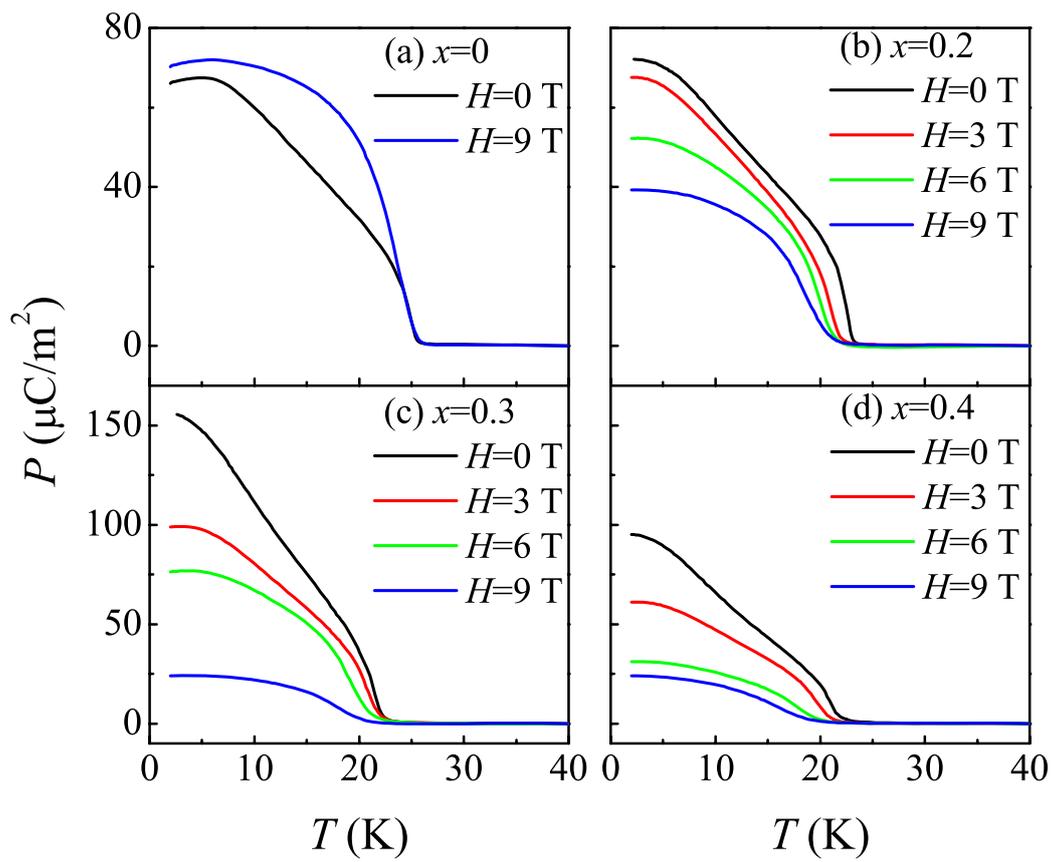

Figure 5

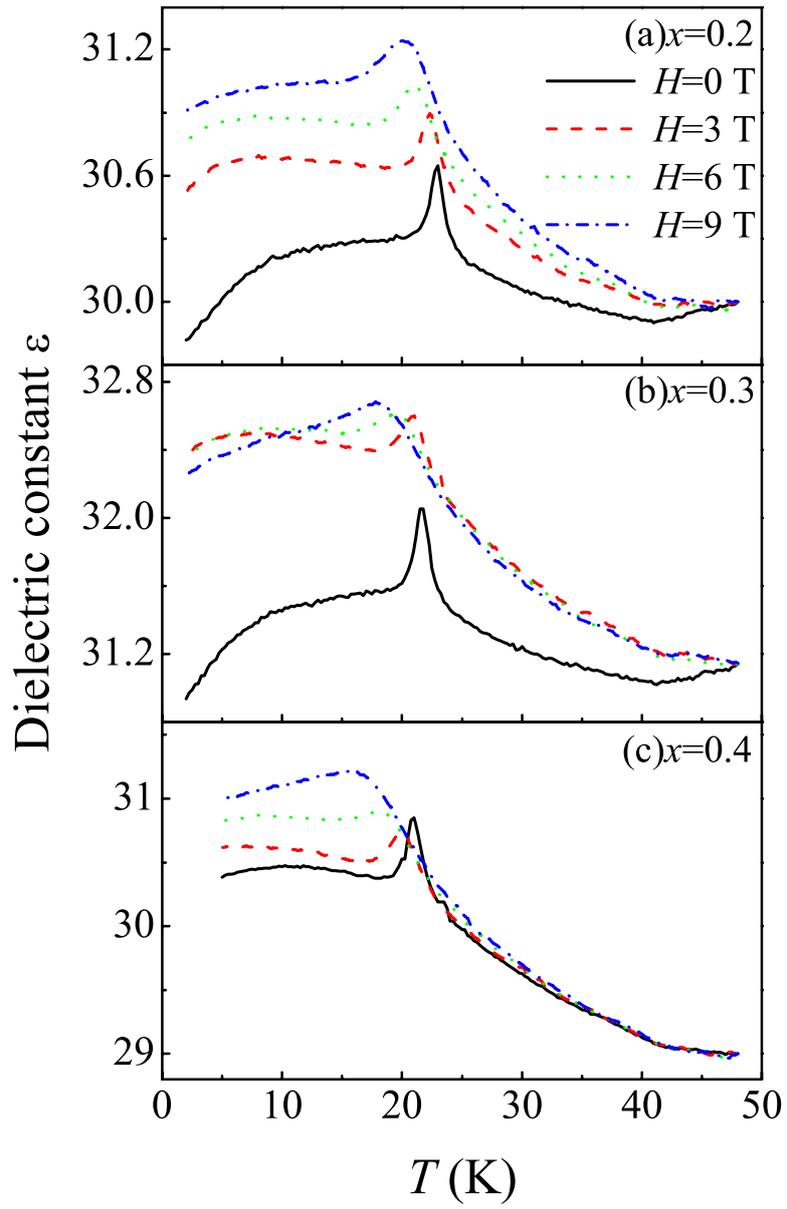

Figure 6

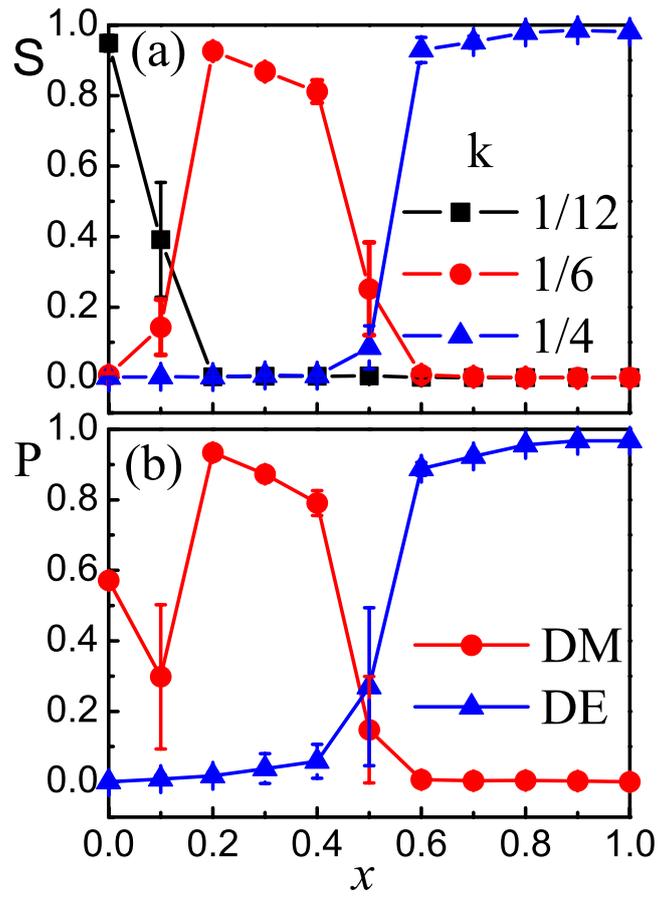

Figure 7

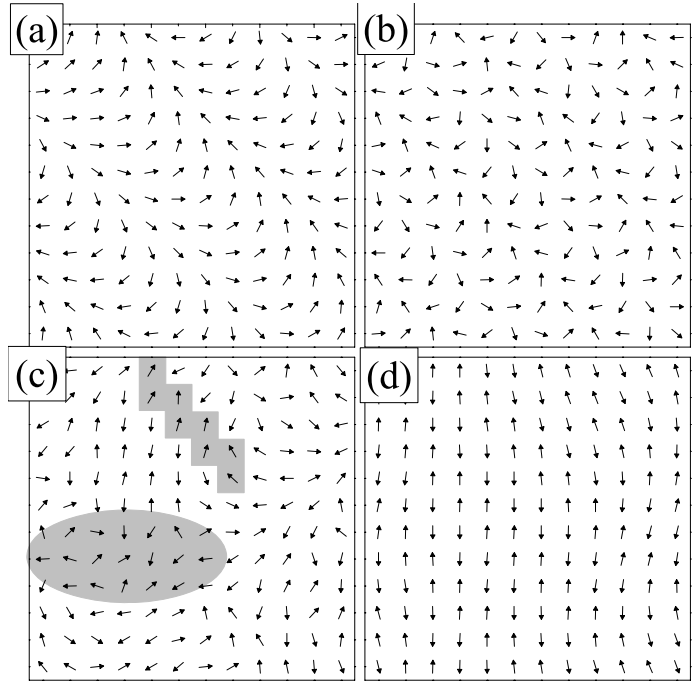

Figure 8